# Recursive introversion, iterative extroversion and transitive ambiversion

Eldar Knar[1]

Institute of Philosophy, Political Science and Religion Studies,
Ministry of Science and Higher Education of the Republic of Kazakhstan

eldarknar@gmail.com
https://orcid.org/0000-0002-7490-8375

**Abstract**

This study proposes an approach to describing personality dynamics through mathematical modelling of introversion, extroversion, and ambiversion processes. Introversion is interpreted as a recursive process characterized by deep self-awareness and inner reflection; extroversion is presented as an iterative process of accumulating and processing external stimuli; and ambiversion is considered transitivity, integrating the interaction of these two opposing processes. The developed model is based on principles of complex systems theory, nonlinear dynamics, and synergetics, enabling the perception of personality as an adaptive and multilayered system.

The results include analytical equations describing the interaction of internal and external factors. Parameters regulating personality dynamics, such as sensitivity coefficients, weighting factors of stimuli, and synergy parameters, are also explored.

The practical significance of the proposed model lies in its application in psychotherapy, education, personnel management, and other areas where it is necessary to consider personality traits and their dynamics. The model provides new opportunities for diagnosing, forecasting, and correcting personality changes, enhancing methods of working with individuals.

The findings open some prospects for quantitative personality analysis, providing a theoretical understanding of its structure and dynamics. They also create a foundation for further interdisciplinary research in psychology, the cognitive sciences, and social systems.

Specifically, the formalization considered in the context of program algorithms allows for the creation of functional procedures for AI and robots with differentiated psychological archetypes (introvert robots, extrovert robots, and ambivert robots).

**Declarations and Statements:**
No conflict of interest
This work was not funded
No competing or financial interests
All data used in the work are in the public domain
Ethics committee approval is not required (without human and animal participation)
Generative AI (LLM or other) was not used in writing the article

---

[1] Fellow of the Royal Asiatic Society of Great Britain and Ireland

# 1. Introduction

Introversion, extroversion, and ambiversion are well-studied and ancient archetypal constructs. However, paradigms are changing. Modern science moves beyond the natural, trivial scientific activities of individuals and transitions into the realm of total formalization of all scientific entities on the basis of quantum models and artificial intelligence. Consequently, old archetypes begin to exist in a new reality, requiring their reevaluation from the perspective of evolving descriptive science into algorithmic and strictly formalized science.

The study of human personality is one of the most complex and multifaceted tasks in psychology, sociology, and philosophy. Traditional personality models, such as the Big Five theory or Jungian typologies, provide valuable insights, but they are often limited by static representations of personality traits and fail to account for the dynamic nature of human behavior and self-awareness. In a rapidly changing world, there is a growing need for more flexible and adaptive personality models that can reflect personality evolution over time and interaction with the environment.

Modern personality research seeks to move beyond traditional typologies and static characteristics, viewing individuals as complex, multidimensional, and dynamic systems. Personality is no longer reduced to a set of constant properties; instead, it emerges as an evolving configuration of processes generating self-awareness, interaction with the external world, and adaptation to changing conditions. In this paradigm, integrating philosophy, psychological theory, mathematics, and complex systems theory becomes particularly significant. This approach allows not only the interpretation of the phenomena of introversion, extroversion, and ambiversion on a deeper level but also the creation of a formal model capable of describing their nonlinear dynamics.

In this study, introversion, extroversion, and ambiversion are presented not as static traits but as fundamentally different yet interconnected dynamic modes of personality functioning. The proposed concept suggests the following:

Introversion is recursion, reflecting in-depth internal analysis and a sequential reinterpretation of one's thoughts and experiences. Introverted processes are represented as recursive schemes where the current state of the personality depends on revisiting its previous states, forming a kind of "internal loop" of self-reflection.

Introversion fundamentally represents a process of self-immersion and reinterpretation of accumulated experiences. It describes the ability of an individual to turn to their past, analyse previous states, and deepen self-awareness. This process can be described through recursion, where the current state of the personality is a function of its previous state. Unlike the static perception of introversion as "reservedness," here, it is interpreted as a mechanism of deepening and repeated rethinking, which is consistent with recursive logic.

Extroversion, on the other hand, is associated with the perception and integration of new external stimuli. It reflects a personality's ability to interact with the external world, accumulate experiences, and respond to environmental changes. In mathematical terms, extroversion is described as an iterative process: each new step adds external influence without deeply altering internal states but rather expands horizons. Iteration forms the basis of extroversion, allowing personality to develop through the accumulation of new experiences.

Extroversion is seen as an iterative process that involves the ability of individuals to expand their experience through the sequential addition of new external data, social interactions, and cognitive and cultural stimuli. Extroverted behavior is interpreted as an iterative process of "accumulating" external information, which acts as the building material for forming new system states.

Ambiversion is understood as transitivity, the ability of the individual to flexibly switch between recursive (introverted) and iterative (extroverted) modes. It acts as a mechanism setting the conditions for transitioning between internal reflection and external expansion, optimizing behavioral strategies depending on the context. Transitivity provides the necessary adaptability, enabling the individual to avoid being stuck in one mode and selectively combining internal and external resources to maximize functional efficiency.

Ambiversion, often viewed as a balance between introversion and extroversion, is treated in this study as a transitive process. It does not merely balance the two poles but actively connects the internal and external aspects of personality, allowing flexible transitions between recursion and iteration. Transitivity ensures system adaptability, enabling the personality to find optimal behavioral strategies depending on the context.

The scientific understanding of personality requires the integration of philosophy, psychology, complex systems theory, and mathematics to create formalized models capable of explaining behavior. This study is based on a philosophical foundation linking internal reflection, external interaction, and adaptive flexibility, manifesting in three key concepts: introversion as recursion, extroversion as iteration, and ambiversion as transitivity.

This perspective transcends classical approaches, which are predominantly descriptive and static. In the proposed paradigm, personality is a nonlinear dynamic system with multiple attractors and transitional states, where introversion, extroversion, and ambiversion are not merely traits but also dynamic patterns emerging from the complex interplay of internal and external factors. From the perspective of complex systems theory, such a model enables the consideration of personality as a self-organizing structure that maintains dynamic equilibrium between competing tendencies for self-immersion and outwards expansion.

In terms of applications, this formalization allows the creation of functional algorithms and procedures for AI and robots with differentiated psychological archetypes (introverted robots, extroverted robots, and ambiverted robots).

From a methodological point of view, relying on mathematical formalizations (using recursive and iterative equations, as well as transitive operators), paves the way for quantitative analysis of personality behavior, enabling the prediction of transitions from one dynamic mode to another under changing conditions. This approach has potential for the development of predictive and optimization models applicable in psychotherapy, coaching, organizational consulting, educational programs, and other practical areas.

The novelty and originality of this study lies in a radical paradigm shift—from describing personality in terms of stable factors to viewing it in terms of dynamic processes. Introversion, extroversion, and ambiversion are no longer considered substantial attributes but rather become operators that transform internal and external experiences. This understanding provides deeper insights into the essence of human psychological life, taking into account the multilevel integration of cognitive, emotional, and sociocultural influences.

Thus, this work lays the philosophical, conceptual, and methodological foundation for developing a new scientific paradigm for studying personality—as a system in which the recursion of introversion, the iteration of extroversion, and the transitivity of ambiversion form a complex and adaptive dynamic picture of self-awareness, behavior, and development.

The introduction of an original philosophy and the development of a formalized model of introversion as recursion, extroversion as iteration, and ambiversion as transitivity open new horizons in personality research. This approach makes it possible to describe complex processes of personality dynamics at a deeper level, creating a theoretical and practical foundation for studying, predicting, and developing human potential.

Of course, psychological archetypes and our formalization are merely imagos (images) of the individual (Jung, 1971). They cannot authentically and adequately reflect the real psychological architecture and construction of personality. However, formalization can provide opportunities for analysing behavioral structures as a first approximation and for modelling in artificial intelligence systems.

2. **Literature Review**

Approaches to studying personality have undergone significant changes over recent decades. From traditional theories, such as Jung's concepts, to contemporary models incorporating complex systems theory and synergetics, each step in personality research has opened new opportunities for deeper understanding.

If we avoid delving too deeply into distant history, Carl Jung (1971) was undoubtedly the first to clearly and systematically classify psychological archetypes (extroversion, introversion, and ambiversion). His groundbreaking ideas laid the foundation for later theories on personality traits and became an essential part of psychoanalysis and cognitive psychology. The studies by Eysenck (1967) and Gray (1981) demonstrated that extroversion and introversion are linked to differences in cortical arousal levels and responsiveness to stimuli. These works emphasize the neurobiological basis of personality traits, which is crucial for their mathematical modelling.

The works of Izard (1991) and Lazarus (1991) highlight the importance of emotions and their relationships with personality traits. These studies provide deeper insights into how emotional regulation affects the balance between introversion and extroversion and personality ability to adapt to changes.

A structured approach to archetypes is addressed in the "Big Five" model (Costa & McCrae, 1992). A sociocultural perspective has been developed in studies investigating the influence of context and environment on psychological archetypes (Markus & Kitayama, 1991). However, these models are static and fail to account for the dynamics of change, necessitating their further development.

Cognitive theories of personality, such as models of parallel information processing (Rumelhart & McClelland, 1986), view personality as a system that processes external and internal stimuli through complex cognitive mechanisms. These approaches enable the development of dynamic personality models that can integrate various factors influencing behavior.

The systemic approach to describing archetypes as dynamic and adaptive phenomena was advanced in formal models by Prigogine (1984). Recursive models are widely used in cognitive psychology to describe memory and learning processes (Rumelhart & McClelland, 1986). Iterative approaches, such as models for processing external stimuli, are used to analyse responses to the environment (Rescorla & Wagner, 1972).

Synergetics, developed by Haken (1983), describes the interaction of internal and external processes in systems, finding application in modelling ambiversion. The concept of transitivity in personality processes involves enhancing adaptability by integrating opposing processes.

Psychotherapeutic practice has long focused on diagnosing personality traits and working with them. Studies in psychotherapy have shown that the use of personality models can improve diagnostics and the development of personalized treatment methods (Beck et al., 1990). The application of mathematical models allows for deeper analysis of personality dynamics.

Biggs (1993) highlights the importance of adapting teaching methods to individual student characteristics, including their tendencies toward introversion or extroversion. Research aimed at identifying students' inclinations toward introversion and extroversion helps optimize teaching methods and improve their effectiveness.

In corporate settings, understanding employees' personality profiles is a critical aspect of enhancing team efficiency and productivity (Goleman, 1995). Mathematical modelling can form the basis for creating more balanced and successful teams. Understanding how different personality types (introverts and extroverts) interact helps create balanced groups that consider their strengths and weaknesses.

The studies of Hofstede (1984) and Triandis (1995) show how introversion and extroversion manifest in different cultures. For example, collectivist cultures (e.g., Japan) emphasize introversion, whereas individualist cultures (e.g., the United States) encourage extroversion.

Thus, individuals can be effectively described via mathematical models that integrate concepts from psychology, the cognitive sciences, and complex systems theory. This study builds on and extends these approaches, proposing new ways to analyse personality dynamics through the interaction of internal and external processes.

**3. Results**

*3.1. Introversion as Recursion*

At one point, Niklas Luhmann defined the attribute of autopoiesis (Maturana, 1980) as follows: "*Autopoiesis is recursive, hence symmetric, and thus a nonhierarchical event*" (Luhmann, 1987). The principles of how basic processes sustain a system's integrity can similarly be applied to specific cases of psychological archetypes, particularly introversion.

Considering the processes of self-immersion, internal analysis, and reinterpretation, which are strongly expressed in introversion, we can interpret introversion as recursion. Indeed, introverts often turn inwards, analysing external environments, internal emotions, and their thoughts. An introvert delves deeper and deeper into layers of consciousness during reflection and analysis. Thus, in psychological processing, the connection between various layers of personal experience and the subconscious is essential for an introvert.

It is evident that an individual does not change instantaneously; their perception and thinking are rooted in prior experiences. Hence, each new psychological or intellectual state of a personality depends on its previous state. While this is also true for extroverts and ambiverts, for introverts, past experiences are dominant in shaping and achieving new states. This recursive dependency on the dynamics of states can be expressed as:

$$I(n-1, D),$$

where:

$n$ - time step (a discrete moment in time when the state is observed),

$D$ - depth level of the state.

The variable n indicates the discrete time index, which describes sequential changes in the state of the individual. For example, $n=0$ can correspond to the initial state of the system (the starting point of analysis or observation), and $n=1,2,3…$ represent subsequent steps describing the evolution of the state.

This reflects the recursion of introversion, where each new state is formed on the basis of the prior state. Overall, the numerical sequence mirrors the process of personality formation, state by state. Suppose that we know only one state of an individual at the initial moment, $I(0,D)$, which is strictly fixed (e.g., 1 for all $D$). Subsequently, at each following step ($n=1$, $n=2$,…), the current state is calculated on the basis of the previous state $I(n-1,D)$, considering internal and external factors.

Here, $D$ represents the depth level of the state, which corresponds to the layered position in an individual's multilayered structure, reflecting the personality's complexity, from superficial to deep layers of self-awareness. Higher values of $D$ correspond to deeper levels of self-awareness, whereas lower values of $D$ reflect more superficial states.

In our interpretation, $D$ is the current level of introversion $I(n,D)$, which interacts with the underlying levels. The deeper the level ($D$), the more it relies on the cumulative influence of all preceding levels. Naturally, all levels are interrelated. Higher levels ($D$) depend on the evolution of lower layers, creating a hierarchical structure where changes at the lower levels significantly impact the higher ones and vice versa.

Superficial levels are associated with sensory perceptions of reality, such as reactions to external stimuli or irritants. In contrast, deeper levels correspond to fundamental aspects of personality, such as self-contemplation, worldview, beliefs, or core memory structures. Changes at these levels occur more slowly.

Thus, the concepts of $n$ and $D$ allow us to describe an individual as a simultaneously evolving (through changes in nnn) and interacting (through changes in $D$) multilayered mental and psychophysical system.

Since individuals do not change instantaneously and rely instead on their past experiences, each new state depends on the previous state. This recursive dependency is expressed as $I(n-1,D)$. However, each new state also carries a proportion of the past state. This principle resembles blockchain technology, where each new block carries specific information about the previous block.

The influence of the previous state can be represented as:

$$(k\,(I(n-1,D))^m),$$

where:

*m>1* - degree of amplification of the influence of past states,

*k* - amplification coefficient determining the "intensity" of the influence of the previous state.

Here, the influence of past states is amplified when the power *m>1*. This enables the modelling of various scenarios where accumulated internal experience (e.g., deep reflection) significantly alters the current state. The coefficient *k*, in a psychological context, might correspond to the depth of introversion or its "energy resource." Clearly, an individual's level of introversion is not a universal constant but varies for different introverts.

The influence of lower levels can be expressed as:

$$w \cdot \sum_{d=1}^{D-1} I(n,d)$$

where the weight coefficient *w* determines the strength of influence that foundational levels exert on the current state. This allows modelling scenarios where the impact of "simpler" processes on higher levels is either significant (*w* is high) or minimal (*w* is low).

This expression can be interpreted as the interconnection of various layers of individual experience, which is consistent with Jung's works and cognitive psychology studies emphasizing the multilayered structure of personality and the interaction of its elements.

Thus, the general equation for introversion as recursion can be written as:

$$I(n,D) = k\,(I(n-1,D))^m + w \cdot \sum_{d=1}^{D-1} I(n,d)$$

This recursive introversion equation is based on the idea of personality's multilayered nature, its capacity for self-reflection, and the influence of external levels on internal states.

### 3.2. **Extroversion as Iteration**

In our notation, extroversion is interpreted as iteration. Broadly speaking, extroversion is understood as the ability of an individual to perceive and accumulate external stimuli through interaction with the surrounding environment. This process is characterized by the linear accumulation of information, meaning that the individual responds to external stimuli sequentially, adding their influence to the current state. Essentially, this is the iteration operation.

In this case, the individual is considered a system that interacts with the external environment. The influence of external factors is integrated iteratively, aligning with approaches from the theory of dynamic systems.

In the first approximation, extroversion is largely determined by external signals and their significance. Each new state of personality is built upon the previous state, taking into account the current external influences. Accordingly, the state of an individual at time *n* depends on their state at the previous time *n−1*:

$$E(n) = E(n-1) + \Delta E,$$

where *ΔE* is the change in state over a single step. This change depends on the external stimuli acting at time n, i.e., during the transition from one state to the next.

The change *ΔE* is determined by the cumulative effect of external factors. Each external stimulus influences the individual's state, with the degree of this influence being regulated by a weight.

This process of change can be formalized as follows:

$$\Delta E\,(n) = s \cdot \sum_{m=1}^{M} cR$$

where
*R* - the value of the external stimulus at time nnn,
*c* - the weight of the influence of the external stimulus, characterizing its significance for the individual,
*s* - the susceptibility coefficient, reflecting the individual's overall sensitivity to external influences,
*M* - the number of factors contributing to state changes.
In its full form, the equation for extroversion as iteration can be written as:

$$E(n) = E(n-1) \ + \ s \cdot \sum_{m=1}^{M} cR$$

The component *E(n−1)* reflects the fact that the individual retains their current state, accumulated from previous steps, highlighting the idea of the linear accumulation of experience. The susceptibility coefficient *s* regulates how strongly the individual responds to external stimuli. Higher values of *s* correspond to higher inverted sensitivity. The summation of the product of the external stimulus and its weight accounts for all the external stimuli acting at time *n*, weighted by their significance. Summation over *M* implies that multiple distinct factors may contribute to the change in state, each adding to the transition.

This equation describes extroversion as a process of accumulating external experience, which is consistent with classical views on the social and cognitive aspects of extroverted behavior. Its structure reflects the key properties of extroversion, such as the linear accumulation of experience, dependence on the external environment, and individual differences in sensitivity to stimuli.

Extroversion is analogous to the process of learning—an individual accumulates external stimuli and updates their state on the basis of their significance and strength.

### 3.3. **Ambiversion as Transitivity**

Ambiversion is not merely a transitional state between introversion and extroversion. It is an independent balanced state characterized by an individual's ability to adapt to internal conditions and external environments, which can be entirely synergistic.

This balanced state, in its simplest representation, can be interpreted through the influence of two factors:
1. Proportional influence of introversion and extroversion. An individual integrates internal analysis through introversion *I(n)* and external influences through extroversion *E(n)*.
2. Synergy of processes. The interaction between introversion and extroversion enhances overall adaptability, creating a nonlinear influence of both factors.

In ambiversion, introversion and extroversion coexist in a specific proportion. Therefore, a proportionality coefficient $\gamma$ can be introduced, which represents the relative weights of introversion and extroversion within the individual. Using $\gamma$, the expression for ambiversion *A(n)* can be written as a transitive function:

$$A(n) = \gamma\, I(n) + (1 - \gamma)E(n)$$

where

$\gamma \in [0,1]$ - proportionality coefficient regulating the dominance of one process over the other.

The coefficient $\gamma$ describes the "tuning" of the individual at a given moment in time. The personality may lean toward introversion, extroversion, or a mixed state. When $\gamma=1$, the individual's state is entirely determined by introversion. When $\gamma=0$, the state is fully defined via extroversion. Intermediate values of $\gamma$ reflect a mixed influence of both processes.

This introvert–extrovert balance is a linear combination of the two processes. Using a distant analogy, this is not the case for Jekyll and Hyde, which exist separately, sequentially, and at different times. Ambiversion is a case where Jekyll and Hyde coexist simultaneously and proportionally within the individual. Ambiversion is not a disorder; it is an inherent property of personality.

The interaction of these two processes (introversion and extroversion) reflects the principles of synergy in complex systems, where the interaction of components leads to new systemic properties.

The synergistic transitivity of ambiversion is interpreted through adaptive behavior arising from the simultaneous interaction of introversion and extroversion. Accordingly, the equation of synergistic transitivity is defined as:

$$T(n) = \delta \cdot \phi \cdot I(n) \cdot E(n)$$

where:

$\delta$ - synergy coefficient describing the strength of interaction between the processes and its impact on the individual's overall state.

$\phi$ - interaction amplification parameter, reflecting the intensity of the mutual influence between introversion and extroversion. This component introduces nonlinearity into the model, accounting for effects such as increased adaptability when both processes are at high levels and a reduced influence of one process if the other is weakened.

By combining the linear balance γ and the transitive component $\delta \cdot \phi$, the final expression for ambiversion *A(n)* becomes:

$$A(n) = \gamma \cdot I(n) + (1 - \gamma) \cdot E(n) + \delta \cdot \phi \cdot I(n) \cdot E(n)$$

In this ambiversion equation, the synergistic component $\delta \cdot \phi$ adds nonlinearity, enhancing the individual's adaptive properties when both processes (introversion and extroversion) are at high levels. For example, when an individual simultaneously analyses internal experiences and responds to external stimuli, the interaction between these processes creates a new quality.

Thus, the equation for ambiversion as transitivity is derived from the a priori assumption that an individual integrates two opposing processes—introversion and extroversion—in a specific proportion and amplifies their interaction through a synergistic effect. It describes not only the basic balance between internal and external processes but also their adaptive interaction, making it suitable for modelling the flexibility and complexity of an individual in changing conditions and in applications that simulate psychological archetypes.

3.4. **Integration**

Combining recursive introversion, iterative extroversion, and transitive ambiversion, the comprehensive model of an individual's psychological archetype can be expressed as:

$$P(n) = I(n) + E(n) + A(n)$$

or more explicitly:

$$P(n) = k\left(I(n-1, D)\right)^m + w \cdot \sum_{d=1}^{D-1} I(n, d) + E(n-1) +$$
$$+ s \cdot \sum_{m=1}^{M} cR + \gamma \cdot I(n) + (1 - \gamma) \cdot E(n) + \delta \cdot \phi \cdot I(n) \cdot E(n)$$

Recursive introversion contributes to exponential growth because of its nonlinear component (*m>1*). The balance between depth (*D*) and the influence of lower levels (*w*) stabilizes the process. External factors ensure the linear accumulation of new experiences, which defines extroversion. The transitive component (*δ·φ*) enhances the interaction between internal and external processes, increasing adaptability.

This model integrates internal and external processes, offering the flexibility and adaptability necessary to understand the complex structure of human behavior and self-awareness. Recursion, iteration, and transitivity serve as metaphors that reflect profound internal processes, external interactions, and their integration.

This model promotes a deeper understanding of personality, adaptability, and growth potential, providing a foundation for further research and practical applications in psychology, sociology, and related sciences. It is particularly relevant for algorithmic implementations of psychological archetypes in artificial neural networks and large language models.

4. **Discussion**

We introduced three key processes: introversion as recursion, extroversion as iteration, and ambiversion as transitivity. These processes were formalized through mathematical expressions, enabling a transition from qualitative description to quantitative analysis.

The presented archetypal model is based on the simple thesis that an individual is a complex dynamic system. Introversion, extroversion, and ambiversion are interpreted and interact through recursive, iterative, and transitive processes, respectively. This approach characterizes the individual not as a mechanical set of static traits but as a system capable of dynamic self-reflection, adaptation, and expansion.

The recursive component *I(n,D)* describes the process of profound self-awareness and reinterpretation of internal experience. The expression for *I(n,D)* enables a quantitative evaluation of how previous states influence the current state, considering the nonlinearity of the process (*m>1*) and the interplay between different levels of self-awareness depth (*w*).

High values of *k* and *m* lead to the exponential growth of introverted processes, reflecting intensive internal analysis and active self-development. In real life, the process of self-awareness and self-reflection often leads to internal deepening and rethinking of past experiences, which frequently results in qualitative changes in one's perception of self and the world.

The nonlinearity of this process illustrates that small changes in perception or behavior can result in significantly larger shifts in personal dynamics.

The iterative component *E(n)* reflects the accumulation of external influences and the expansion of external interactions. The formula for *E(n)* demonstrates how external factors and their weights affect the state of extroversion. The susceptibility coefficient regulates the degree of an individual's reaction to external stimuli.

This is analogous to the linear accumulation of external experience, enabling individuals to broaden their horizons through sequential steps of interaction with the external ecosystem.

The iterative nature of extroversion signifies a linear process of accumulating new external data, which nonetheless plays a vital role in adapting the individual to changing environmental conditions. In a psychological context, this aligns with the concept of social adaptation, where a person accumulates new social interactions, experiences, and impressions.

Ambiversion, represented by *A(n),* integrates both internal (introversion) and external (extroversion) processes, balancing their influence through the proportionality coefficient γ and amplifying their interaction via the synergistic term $\delta \cdot \phi$. This transitive component ensures adaptability and creates new systemic properties that emerge from the interaction of opposing forces.

The combined model reflects the dynamic nature of personality and provides a mathematical foundation for studying the flexibility and adaptability of individuals. Its potential applications extend to psychology, artificial intelligence, and the simulation of human-like behavior in technological systems.

The ambivert component *A(n)* is interpreted through a transitive function integrating introversion and extroversion. The parameters γ and $\delta \cdot \phi$ regulate the balance between introversion, extroversion, and their synergy. This component provides flexibility, allowing the individual to adapt to changing conditions and find an optimal equilibrium between internal analysis and external interactions.

In a psychological context, this corresponds to a higher level of adaptability and the ability of an individual to balance various aspects of their existence.

High values of $\gamma$ (closer to 1) - the psychological archetype leans toward introversion, which may result in reduced external interactions and increased inwards focus. Low values of $\gamma$ (closer to 0) - extrovert processes dominate, promoting active engagement with the external ecosystem. The transitivity parameter $\delta \cdot \phi$ reflects the individual's dynamic adaptability, enabling transitions between introverted and extroverted modes depending on the current conditions. High transitivity values correspond to flexible transitions, preventing extremes and maintaining stable balance.

The presence of nonlinearity ($m>1$) in the recursive component of introversion introduces complex dynamic effects into the model. This allows the psychological archetype system to exhibit exponential growth of introverted processes under stable external conditions, corresponding to intensive internal analysis and self-development.

The iterative component of extroversion $E(n)$ interprets the linear accumulation of external influences. This means that individuals can expand their experience and interactions with the external ecosystem.

The combined effect of recursion and iteration, mediated by the transitive component, generates complex dynamic patterns. This includes the possibility of attractors and chaotic regimes under certain parameter combinations. Psychological or archetypal attractors represent a separate topic; understanding how trajectories of archetypes change or attract allows exploration of more complex interactions between recursive, iterative, and transitive properties of psychological archetypes.

The proposed model significantly differs from traditional approaches to studying personality, such as Big Five theory or Jung's typologies, which focus on static traits. In contrast, this model views personality as a dynamic system capable of change and adaptation through the interaction of internal and external processes.

It enables an interpretation of an individual's dynamism, describing changes over time by considering both internal reflective processes and external stimuli. Additionally, the model integrates processes by unifying introversion, extroversion, and ambiversion into a single system, demonstrating their interconnections and mutual influences.

Overall, this model provides a quantitative description of individual dynamics as a first approximation, allowing predictive analysis and process optimization for personality development.

Naturally, this model is relatively primitive and thus limited. More precise definitions and parameter adjustments based on large empirical datasets are needed.

A rigorous validation process is needed, relying on real-world data obtained from long-term studies and comprehensive psychological measurements. The development of recursive, iterative, and ambitious computational models for interpreting psychological archetypes is a complex task.

Hypothetical practical applications:

Personal balance assessment and correction—The model can be used to assess and adjust an individual's personality balance. For example, a client with dominant introversion might focus on developing social skills. A person with excessive extroversion could work to enhance self-reflection.

Educational environment—the model could help design individual learning strategies tailored to the dominant processes of a student's personality. For example, introverted students might benefit more from independent projects. Extroverted students might perform better in group assignments.

The corporate environment model could assist in evaluating employees' personality profiles and assigning optimal roles within teams. For example, introverts could focus on analytical tasks. Extroverts could excel in client-facing roles.

This model, which offers a mathematical description of personality dynamics through recursion, iteration, and transitivity, provides promising tools for analysing, predicting, and optimizing personality changes. Unlike traditional static models, it dynamically describes personality as a self-developing system capable of deep adaptation and flexibility.

Despite existing limitations, the model has significant potential for scientific research into archetype algorithmization in AI environments and practical applications in psychotherapy, coaching, education, and personnel management.

Future studies aimed at empirical validation and model expansion will make this method even more universal and precise.

**5. Conclusion**

In conclusion, we summarize the findings and implications of this study:
- The models are based on comprehensive theoretical concepts from psychology, complex systems theory, and the cognitive sciences. Introversion, extroversion, and ambiversion have been translated from qualitative concepts into quantitative parameters through logical algorithms, enabling the study of their interactions and mutual influences.
- The developed models of introversion as recursion, extroversion as iteration, and ambiversion as transitivity represent tools, algorithms, and paradigms for describing individual dynamics. Introversion is not merely a tendency toward inner reflection but also a dynamic process of deepening experience.
- Introversion, interpreted as a recursive process, describes profound inner reflection that is dependent on accumulated experience.

- Extroversion is formalized as an iterative accumulation of external influences, and extroversion represents the sequential interaction of an individual with the external environment. It involves not only social activity but also systematic interaction with external stimuli.
- Ambiversion is mathematically expressed as a synergistic interaction of these processes, and ambiversion provides flexibility and adaptability in changing conditions. It is not a static intermediate state but an active integration of two processes.
- The model accounts for nonlinear effects, such as the exponential influence of internal experience ($m>1$) or the synergistic effect of introversion–extroversion interactions ($\delta \cdot \phi$).
- Interaction between depth levels ($D$) ensures multilayered dynamics, reflecting the complex nature of personality.
- The introduction of parameters $\gamma, \delta, \phi, w$ and $s$ allows the model to adapt to a wide range of individual profiles and situations.
- The model is capable of approximating extreme states (pure introversion or extroversion) as well as the balanced state of ambiversion.
- The model can be used for diagnosing and adjusting personality balance. For example, individuals with dominant introversion can be guided to utilize more external resources, and extroverts can focus on developing internal reflection.
- The model supports the adaptation of teaching methods to individual characteristics, optimizing the learning process by accounting for personal predispositions; introverted students may benefit more from independent projects, and extroverted students might thrive in group-based assignments.
- The model allows for assessing employee personality profiles and selecting optimal roles in teams. For example, introverts might excel in analytical tasks, and extroverts could succeed in client-facing roles.
- Practical application of the model requires validation with real-world data. Research and testing using psychological methods are necessary.
- The model does not yet account for emotional, cognitive, and cultural factors, which can significantly impact personality.
- Research is needed to calibrate model parameters on the basis of data from psychological testing.
- The model can be extended to include additional processes, such as emotional dynamics, cognitive aspects, or the influence of social factors.
- Creating software for visualizing and analysing personality dynamics on the basis of this model is essential. Additionally, the model can be expanded to describe group and collective dynamics.
- Models based on recursion, iteration, and transitivity describe personality as a dynamic and multilayered system where internal and external processes interact and evolve over time.

- The full, expanded model integrates introversion, extroversion, and ambiversion into a single dynamic system capable of describing a wide spectrum of personality states, from introversion to full social activity, with adaptive mechanisms.
- The formalization presented here can be implemented in programmatic algorithms, enabling the creation of functional procedures for AI and robots with differentiated psychological archetypes (introvert robots, extrovert robots, and ambivert robots).